\def \s{~\rm{s}}
\def \km{~\rm{km}}

\def \K{~\rm{K}}

\def \erg{~\rm{erg}}

\def \yr{~\rm{yr}}

\def \kpc{~\rm{kpc}}

\documentclass[12pt,preprint]{aastex}

\usepackage{natbib}
\usepackage{epsfig}
\begin{document}

\title{MAGNETIC ACTIVITY IN STELLAR MERGER PRODUCTS}

\author{Noam Soker\altaffilmark{1}
and Romuald Tylenda\altaffilmark{2}
}

\altaffiltext{1}{Department of Physics, Technion-Israel Institute of
Technology,  32000 Haifa, Israel;
soker@physics.technion.ac.il.}

\altaffiltext{2}{Department for Astrophysics, N.Copernicus Astronomical Center,
Rabia\'nska 8, 87-100 Toru\'n, Poland;
tylenda@ncac.torun.pl}

\begin{abstract}

We study the expected X-ray luminosity of stellar merger products several
years after merger.
The X-ray emission is assumed to result from magnetic activity.
The extended envelope of the merger product possesses a large convective
region and it is expected to rotate fast.
The rotation and convection might give rise to an efficient dynamo operation,
therefore we expect strong magnetic activity.
Using well known relations connecting magnetic activity and X-ray luminosity
in other types of magnetically active stars, we estimate that the strong
X-ray luminosity will start several years after merger,
will reach a maximum of  $L_x \sim 3 \times 10^{30} \erg \s^{-1}$, and
will slowly decline on a time scale of $\sim 100 \yr$.
We predict that X-ray emission from V838 Mon which erupted in 2002
will be detected in 2008 with 20 hours of observation.

\end{abstract}


\section{INTRODUCTION}
\label{intro}


The eruption of V838~Mon in 2002 (Brown 2002) and subsequent studies of its
observed evolution (Munari et~al. 2002; Kimeswenger et~al. 2002; Crause et~al.
2003; Kipper et~al. 2004; Tylenda 2005), as well as, of other similar objects,
i.e. V4332 Sgr (Martini et~al. 1999; Tylenda et~al., 2005) and M31~RV
(Mould et~al. 1990) have led to suggestions that these observed events were
likely to be due to stellar mergers (Soker \& Tylenda 2003, Tylenda \& Soker 2006).
Soker \& Tylenda (2006), who termed these events mergebursts, discuss the
different channels to produce a mergeburst.

For hours to months after merger the merger product is very luminous
(e.g., Soker \& Tylenda 2003; Bally \& Zinnecker 2005; Tylenda 2005).
For a grazing collision (namely, not a head on collision) an extended
envelope is inflated by the merging stars.
Still on a longer time scale, the mass loss process, both mass loss rate
and geometry, is strongly influenced by the merger event (e.g.
Morris \& Podsiadlowski 2006).
On a much later time of hundreds of years and longer, after the merger
products reaches equilibrium, the process can alter
the evolution of the star on the HR diagram (e.g., Podsiadlowski et al. 1990), like the
formation of blue stragglers (e.g. De Marco et al. 2005; Sills et al. 2005).

In the present study we examine whether and when the merger product can
become magnetically active, a process that might be observed in the X-ray and
radio bands.

\section{THE EXPECTED MAGNETIC ACTIVITY}
\label{model}

\subsection{The Rossby Number and X-ray Luminosity}

The parameter that best indicates the level of magnetic activity of main
sequence stars (e.g., Pizzolato et al. 2003), pre-main sequence stars
(e.g., Preibisch et al. 2005), and subgiants (or G giants; e.g., Gondoin 2005)
is the Rossby number
\begin{equation}
Ro \equiv \frac{P_{\rm rot}}{\tau_{c-b}}=\frac{P_{\rm rot}}{\alpha H_{p-c}/v_{c-b}},
\label{rossby}
\end{equation}
where $P_{\rm rot}$ is the rotation period of the star, $\tau_c=\alpha H_p/v_c$ is
the convection overturn time, $H_p$ is the pressure scale hight, $\alpha H_p$ is
the mixing length, and $v_c$ is the velocity of the convective cells.
In particular, the correlations of some properties of the magnetic activity in main
sequence stars with the Rosbby number and the explanation of these in the frame of
the $\alpha \omega$ dynamo model are well established
(e.g., Brandenburg et al. 1988; Saar \& Brandenburg 1999).
The subscript `{\it b}' in equation (\ref{rossby})
indicates that the value of $\tau_c$ is calculated at the
bottom (inner boundary) of the envelope convective region, or just above it.
In stars having a fully convective envelope, e.g., low mass main sequence stars,
it is complicated to calculate $\tau_{c-b}$. In that case one can define
the global overturn time
\begin{equation}
\tau_{c-{\rm global}} \equiv \int^{R_\ast}_{R_{b}} \frac{dr}{v_c},
\label{global}
\end{equation}
where $R_\ast$ is the stellar radius and $R_b$ is the radius at the bottom
of the envelope convective region.
Kim \& Demarque (1996) find for main sequence stars the relation
$\tau_{c-b} \simeq 0.5 \tau_{c-{\rm global}}$.

We are interested in the x-ray emission resulting from magnetic activity.
The magnetic flux on the surface of magnetically active main sequence stars
is proportional to the X-ray luminosity $L_x$ (e.g., Pevtsov et al. 2003).
The ratio of the X-ray luminosity to the bolometric luminosity of
main sequence stars has a general relation of
\begin{equation}
\frac {L_x}{L_{\rm bol}} = C_x Ro^{-2}; \qquad  0.15 \la Ro \la 10.
\label{lxlb}
\end{equation}
${L_x}/{L_{\rm bol}}$ saturates at a values of $\sim 10^{-3}$
for $Ro \la 0.15$, while no activity is detected for $Ro \ga 10$
(Pizzolato et al. 2003).
For main sequence stars $C_x \simeq 10^{-5}$ (Pizzolato et al. 2003),
while for G giants (subgiants) $C_x \sim 10^{-6}$ (Gondoin 2005).
YSOs are usually in the saturated regime, and show higher activity than main
sequence stars with the same mass or bolometric luminosity (Preibisch et al. 2005).

\subsection{The Rossby Number in Inflated Merger Products}

Following Tylenda \& Soker (2006) we assume that the merger remnant is
composed of a more or less undisturbed pre-merger primary star
of mass, $M_1$, and radius,
$R_1$, surrounded be an envelope of mass, $M_{\rm env}$, inflated up to an
outer radius, $R_{\rm env}$.

Merger products are expected to contract more or less
along the Hayashi line (Tylenda 2005; Tylenda \& Soker 2006).
However, they are different from young stellar objects
(YSOs) contracting along the Hayashi line. An inflated merger product has
a well defined and relaxed central region$-$the pre-merger primary star,
while the contracting envelope contains a relatively small amount of mass.

In that respect, the inflated merger remnants are more similar to late
asymptotic giant branch (AGB) and post-AGB
stars; both classes of objects share the following properties:
\begin{enumerate}
\item  Radius of tens to hundreds solar radii.
\item Luminosity of $\sim 3\times 10^3-10^5 L_\odot$.
\item Cool envelope, $T_{\rm eff} < 10^4 \K$.
\item Extended convection region in the envelope. To compensate for the low
density in the expression for convective energy transport,
the convective velocity must be large.
\item A low mass envelope with a compact massive center: the stellar core in late
AGB stars and post AGB stars, and the primary in inflated merger products.
\end{enumerate}

Based on these properties, we proceed as follows.
To estimate the convective velocity $v_c$ we use results of late AGB
and post-AGB stars (Soker \& Harpaz 1992; 1999).
These results show that just below the photosphere, where the temperature is
$T \sim 10^4 \K$, the convection velocity is $v_c \sim 8 \km\s^{-1}$.
In the stellar numerical code the convection velocity is limited by the
isothermal sound speed, because for higher convection velocities the dissipation
is large, and the convective cells rapidly slow down.
The value of $v_c$ stays at $v_c \simeq 8-20 \km \s^{-1}$ in most of the
envelope.
We will therefore take $v_c=10 \km \s^{-1}$, and use the global
convective overturn time as defined in equation (\ref{global}).
Using $\tau_{c-b} \simeq 0.5 \tau_{c-{\rm global}}$ (Kim \& Demarque 1996),
and $R_b \ll R_\ast$, we take for merger remnants
\begin{equation}
\tau_{c-m} \simeq 0.5\frac{R_{\rm env}}{v_c} \simeq 40    
\frac{R_{\rm env}}{100 R_\odot}{\rm days}.
\label{taucm}
\end{equation}

A similar result is obtained if we consider, following Tylenda (2005),
the envelope of the merger product to be an $n=3/2$ polytrope,
and calculate $\tau_c$ at the middle of the envelope $R=R_{\rm env}/2$.
In an $n=3/2$ envelope the pressure scale hight has its maximum value
of $H_p \simeq R_{\rm env}/10$ at the middle of the envelope.
Taking for the ratio of mixing length to pressure scale hight
$\alpha=1.86$ (Kim \& Demarque 1996) would give
$\tau_{c-m} \simeq 15 (R_{\rm env}/100 R_\odot)$~days.
On the other hand, our estimate of $v_c$ might be too large,
with an underestimate of $\tau_c$, as pre-main sequence stars
have $\tau_c \simeq 200$~day (Preibisch et al. 2005).

The inflated envelope of the merger remnant stores an angular
momentum comparable to that of the pre-merger orbital motion of the secondary.
For an $n=3/2$ polytropic envelope having $R_{\rm env} \gg R_1$ the
moment of inertia can be approximated as
$I \simeq 0.11 M_{\rm env} R_{\rm env}^2$.
We assume that after several dynamical time scales the convection in the envelope
brings the envelope to a solid body rotation.
Assuming that the secondary had a Keplerian velocity as it collided with the
primary at radius $R_1$ and that the merger product envelope has a mass
comparable to that of the secondary, we can estimate a rotation period of
the envelope as
\begin{equation}
P_{\rm rot} \simeq 130             
\left( \frac{R_{\rm env}}{100 R_\odot} \right)^2
\left( \frac{M_1}{M_\odot} \right)^{-1/2}
\left( \frac{R_1}{R_\odot} \right)^{-1/2} {\rm days}.
\label{prot1}
\end{equation}
Equivalently we can define a parameter $\eta$ being the ratio of the
envelope rotation velocity to the Keplerian velocity $v_{\rm Kep}$(or Keplerian
period $P_{\rm Kep}$ to rotation period) at $R_{\rm env}$,
namely
\begin{equation}
\eta \equiv \left(  \frac{v_{\rm rot}}{v_{\rm Kep}} \right)_{R_{\rm env}}
\simeq 0.9 \left( \frac {100 R_1}{R_{\rm env}} \right)^{1/2}.
\label{eta}
\end{equation}
The second equality uses equation (\ref{prot1}).

As it is clear from the above equations, when the remnant contracts, it spins-up.
We assume that after it reaches a rotation velocity of
some fraction $\eta_{\rm max}$ of its break-up (Keplerian) velocity
mass loss keeps the value of $\eta$ unchanged.
When it happens, the rotation period is
\begin{equation}
P_{\rm rot} \simeq 230                
\left( \frac{\eta_{\rm max}}{0.5} \right)^{-1}
\left( \frac{R_{\rm env}}{100 R_\odot} \right)^{3/2}
\left( \frac{M_1}{M_\odot} \right)^{-1/2}{\rm days},
\label{prot2}
\end{equation}

The Rossby number (eq. \ref{rossby}) for the inflated merger remnant can be obtained from
equation (\ref{taucm}) using equations (\ref{prot1}) or (\ref{prot2}), i.e.
\begin{equation}
Ro({\rm merger}) \simeq 3     
\left( \frac{R_{\rm env}}{100 R_\odot} \right)
\left( \frac{M_1}{M_\odot} \right)^{-1/2}
\left( \frac{R_1}{R_\odot} \right)^{-1/2}
\label{rom1}
\end{equation}
if equation (\ref{eta}) gives $\eta < \eta_{\rm max}$ or
\begin{equation}
Ro({\rm merger}) \simeq 6   
\left( \frac{\eta_{\rm max}}{0.5} \right)^{-1}
\left( \frac{R_{\rm env}}{100 R_\odot} \right)^{1/2}
\left( \frac{M_1}{M_\odot} \right)^{-1/2}.
\label{rom2}
\end{equation}
otherwise.

\subsection{The X-Ray Luminosity of Inflated Merger Products}

As the merger products are somewhat similar to giant stars, we should
take $C_x=10^{-6}$ in equation (\ref{lxlb}) (Gondoin 2005).
The operation of an $\alpha \omega$ dynamo in the envelope of AGB stars that
were spun-up  by low mass companions spiraling inside their envelope
was considered before (Nordhaus \& Blackman 2006 and references therein).
However, AGB stars that are expected to rotate very slowly and have large
Rossby number $Ro \gg 10$ (Soker \& Zoabi 2002), do amplify magnetic fields,
as evidenced by polarization of maser emission in local regions around these stars
(Szymczak 1998; Vlemmings 2005).
In seems as if a dynamo based mainly on convection, and not on convection+rotation
(the $\alpha \Omega$ dynamo model), can also amplify magnetic fields in giants
(Soker \& Zoabi 2002; Soker \& Kastner 2003; Dorch 2004), but
not as efficiently as the $\alpha \omega$ dynamo we appeal to here.
Therefore, although our envelope model is similar to that of AGB stars,
the dynamo model we use is much more efficient than that expected in
AGB stars.
By taking $C_x=10^{-6}$ we might underestimate the
X-ray luminosity of merger remnants.
Using equation (\ref{rom1}) or (\ref{rom2}) in equation (\ref{lxlb})
with $C_x=10^{-6}$
we find the expected X-ray luminosity of the contracting envelope
\begin{equation}
L_x \simeq 4 \times 10^{30}
\left( \frac{R_{\rm env}}{100 R_\odot} \right)^{-2}
\left( \frac{M_1}{M_\odot} \right)
\left( \frac{R_1}{R_\odot} \right)
\left( \frac{L_{\rm bol}}{10^4 L_\odot} \right)
\erg \s^{-1} ,
\label{lxm1}
\end{equation}
if equation (\ref{eta}) gives $\eta < \eta_{\rm max}$, or
\begin{equation}
L_x \simeq 1.2 \times 10^{30}
\left( \frac{\eta_{\rm max}}{0.5} \right)^{2}
\left( \frac{R_{\rm env}}{100 R_\odot} \right)^{-1}
\left( \frac{M_1}{M_\odot} \right)
\left( \frac{L_{\rm bol}}{10^4 L_\odot} \right)
\erg \s^{-1},
\label{lxm2}
\end{equation}
otherwise.

\section{RESULTS FOR V838 Mon}
\label{v838}

We can apply the general derivation of the previous section to predict the
expected evolution of the X-ray luminosity of V838 Mon.

\begin{figure}
{\includegraphics[scale=0.88]{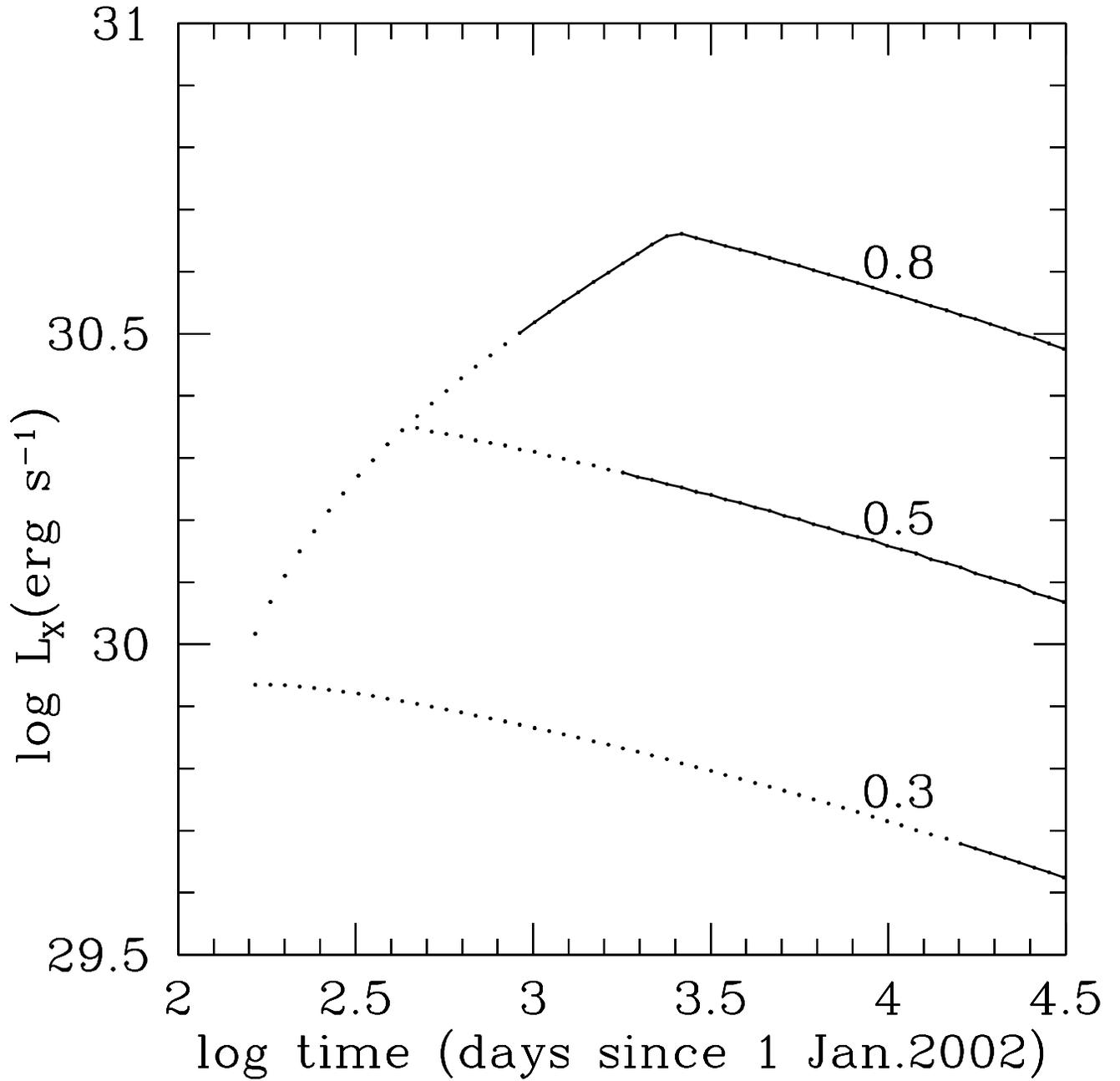}}
\caption{The expected evolution of the X-ray luminosity from the
magnetically active V838~Mon remnant according to the merger model.
The curves are labelled with the value of $\eta_{\rm max}$,
the maximum ratio of rotation velocity to Keplerian velocity on the equator.
Dotted parts indicate the time period when the Rossby number is $Ro > 5$,
where the $\alpha \omega$ dynamo model is less efficient.}
\label{lumx}
\end{figure}

As discussed in Tylenda (2005) the observed decline in flux of V838 Mon after its
eruption can be
well described by gravitational contraction of a low-mass inflated envelope
sitting on top of an early B-type main sequence star. Assuming more recent
determinations of the distance to V838~Mon giving a value of $\sim 6$~kpc
(Sparks et~al. 2007; Bond \& Afsar 2007) (compared to 8~kpc assumed in Tylenda 2005)
the parameters of the model fitted to the observed decline become:
$M_1 \simeq 7 M_\odot$,
$R_1 \simeq 3.5 R_\odot$ and $M_{\rm env} \simeq 0.12 M_\odot$. At the
beginning of the contraction (August-September 2002) the envelope radius was
$R_{\rm env} \simeq 2000 R_\odot$. Using the same approach as in Tylenda
(2005) and the above parameters, we can follow the contraction of
the V838~Mon remnant to obtain the evolution of $R_{\rm env}$ and
$L_{\rm bol}$ with time. This allows us to predict from the relations derived
in Section \ref{model} the evolution of the X-ray luminosity. The
results are presented in Fig.~\ref{lumx}.
The curves show the evolution of the X-ray luminosity with time and are
labelled with the value of $\eta_{\rm max}$ used when caluculating the
luminosity from equations (\ref{lxm1}) and (\ref{lxm2}).
The rising parts of the two upper curves correspond to the initial,
constant angular momentum phase of the remnant contraction ($\eta < \eta_{\rm max}$).
The declining
parts show the phase when the envelope is loosing angular momentum via mass loss so
that the condition $\eta = \eta_{\rm max}$ is kept. The dotted parts of the
curves show the periods when the Rossby number is greater then 5. We expect
that during this period the dynamo is less effective than assumed in our
estimates so our results may overestimate the X-ray luminosity.

With {\it Chandra} 23 hours of observations of the Orion Nebula,
at a distance of $0.45 \kpc$, Feigelson et al. (2002) could detect
sources with luminosity down to $L_x=10^{28} \erg \s^{-1}$.
For a distance of $6 \kpc$ to V838 Mon (Sparks et~al. 2007; Bond \& Afsar 2007),
and with a similarly long observation, we expect to detect any emission if
$L_x \ga 2 \times 10^{30} \erg \s^{-1}$.
We conservatively took $C_x=10^{-6}$ in equation (\ref{lxlb}),
as appropriate for subgiants (Gondoin 2005) rather than
$C_x=10^{-5}$ as appropriate for main sequence stars (Pizzolato et al. 2003).
More than that, bright pre main sequence stars with no accretion disk
are X-ray brighter than those with disks (Preibisch et al. 2005).
As V838 Mon does not have an accretion disk, it is quite possible
that we underestimate the X-ray luminosity of merger products in
equations (\ref{lxm1}) and (\ref{lxm2}) by up to an order of magnitude.
Therefore, it is quite possible that 10 hours of XMM-Newton or {\it Chandra}
observation could detect X-rays from V838~Mon at present and in coming
years.

V838 Mon was observed with {\it Chandra} for $6800\s$ a year after its outburst
by Orio et~al. (2003) who were able to put only an upper limit of
$F_X \le 6.5 \times 10^{-14} {\rm erg}\,{\rm cm}^{-2}\,{\rm s}^{-1}$.
With a distance of $\sim 6$~kpc this corresponds to $L_X \le 2.8
\times 10^{32} {\rm erg}\,{\rm s}^{-1}$ which is well above our predictions.

\section{SUMMARY}
\label{summary}

According to the stellar merger model of the V838 Mon outburst
and similar merger products (which we term mergebursts),
a large envelope is formed around the more massive of
the two merging stars.
The envelope then contracts on a thermal time scale.
The merger remnant should become a fast rotator as it contracts.
As the remnant contracts more or less along the Hayshi line,
its envelope possesses a large convective region.
The fast rotation and the envelope convection are the two ingredients
required in the $\alpha \omega$ dynamo model$-$a successful model
for magnetic activity of main sequence stars, pre-main sequence stars,
and subgiants.

We applied the $\alpha \omega$ model to contracting merger products
by using the Rossby number (eq. \ref{rossby}), and the relation between
the Rossby number and X-ray luminosity known for magnetically active
stars, scaled according to the expression for subgiant
(or G giant) stars (eq. \ref{lxlb}).
We also assumed that after the contracting product reaches
some fraction $\eta_{\rm max}$ of its break-up (Keplerian) velocity,
this ratio does not increase any more, because a stellar wind removes angular
momentum from the envelope.
Our final (and conservative) prediction for the X-ray luminosity of magnetically active
merger products are given by equation (\ref{lxm1}) for merger products
before they reach our assumed maximum rotation rate, and by equation
(\ref{lxm2}) for merger products rotating at $\eta_{\rm max}$.

In section \ref{v838} we apply  the results to our model of V838 Mon.
The results are presented in Fig. \ref{lumx} for three values of the assumed
maximum rotating rate $\eta_{\rm max}$, as marked near the lines.
For too large Rossby numbers $Ro \ga 10$ (Pizzolato et al. 2003;
we here take a stronger constraint of $Ro \ga 5$) of  the $\alpha \omega$ dynamo
is not efficient any more.
The dotted lines are the evolutionary stages where the expected Rossby
number of V838 Mon is $Ro>5$, and we expect no strong  magnetic activity.

>From Fig. \ref{lumx} we learn the following.
\begin{enumerate}
\item  There is no magnetic activity at the first several years,
and hence no X-ray emission is expected.
The observation by Orio et al. (2003) was made a year after the
outburst, when no magnetic activity and no X-ray emission is expected.
\item For a reasonable values of maximum rotation rate
$0.4 \la \eta_{\rm max} \la 0.8$, V838 Mon will reach a maximum activity
at 6-8 years after outburst. The expected
X-ray luminosity then slowly declines.
\item The X-ray luminosity in the coming years will be
$L_x \sim 3 \times 10^{30} \erg \s^{-1}$. At the distance of V838 Mon the
expected X-ray flux is
$F_X \sim 6 \times 10^{-16} {\rm erg}\,{\rm cm}^{-2}\,{\rm s}^{-1}$.
We estimate that with 100,000 seconds of observation this emission
can be detected.
\end{enumerate}

We therefore highly encourage 100,000 seconds of X-ray observation
of V838 Mon in 2008.
Even if no X-ray is detected, the results is of some importance,
as it can strongly constrain models for V838 Mon, e.g., rules out
accreting white dwarf.
Orio et al. (2003) noted that their null detection rules our a symbiotic-like
event to the V838 Mon outburst.

We point out that the null detection of X-ray emission from two AGB stars
(Kastner \& Soker 2004) is not directly relevant to the case of V838 Mon.
First, and most important, our prediction is based on the $\alpha \omega$
dynamo model, namely, the amplification of the magnetic field by the operation
of both rotation and convection, which is known to be very efficient.
On the other hand, predictions for AGB stars are based on the
amplification of the magnetic field by convection alone (Soker \& Zoabi 2002),
which is thought to be much less efficient.
Second, V838 Mon is an order of magnitude more massive than
an upper AGB star. We predict the magnetic activity to take place when
the radius, luminosity and temperature of V838 Mon is similar to that
of upper AGB star.
Due to the higher mass we expect the mass loss rate to be smaller, and the
wind speed to be faster. Therefore, the column density to the expected
X-ray emitting region will be much lower.

Finally, the magnetic fields might be detected also in masers spots.
Deguchi (2005) and Claussen (2005) report the detection of SiO maser around v838 Mon.
We predict that if maser emission, in SiO, H$_2$O, or OH, will be observed
from 2007, some regions might show polarization indicating the presence of
magnetic fields, similar to the case around AGB stars, e.g., Vlemmings et al. (2005).

\acknowledgments
We thank an anonymous referee for useful comments.
This research was supported in part by the Asher
Fund for Space Research at the Technion, as well as, from the Polish State
Committee for Scientific Research grant no. 2~P03D~002~25.

\end{document}